\documentclass{article}
\usepackage{calc}
\usepackage{graphicx}
\usepackage{here}
\usepackage{color}
\usepackage{amsmath}
\usepackage{longtable}
\usepackage{graphics}
\usepackage{subfigure}
\usepackage{setspace}
\usepackage{units}
\usepackage[margin=1in]{geometry}
\usepackage{dcolumn}
\usepackage{bm}
\graphicspath{{figures/}}

\def\g2{{$(g-2)$} }
\def\be{\begin{equation}}
\def\ee{\end{equation}}
\def\ben{\begin{enumerate}}
\def\een{\end{enumerate}}
\def\bi{\begin{itemize}}
\def\ei{\end{itemize}}
\def\bs{\begin{slide}}
\def\es{\end{slide}}
\def\bea{\begin{eqnarray} }
\def\eea{\end{eqnarray} }
\def\bc{\begin{center} }
\def\ec{\end{center} }

\newcommand{\amu}[1][]{\ensuremath{a_{\mu^{#1}}}}

\begin{document}

\title{ The Muon $(g-2)$ Spin Equations, the Magic $\gamma$,\\
 What's small and  what's not.}

\author{ 
James P. Miller$^\dag$ and B. Lee Roberts$^\ddag$ \\
Department of Physics, Boston University\\ 
\vspace{0.05in}
Boston, MA, USA 02215\\
 $^\dag$miller@bu.edu\\
 $^\ddag$roberts@bu.edu
}


\maketitle

\singlespacing 

{\normalsize                                                          

\begin{abstract}
We review the spin equations for the muon in the 1.45~T  muon \g2 storage
ring, now relocated to 
 Fermilab. Muons are stored in a uniform 1.45~T magnetic field, and vertical
 focusing is provided by four sets of electrostatic
 quadrupoles placed symmetrically
 around the storage ring. The storage ring is operated at a Lorentz factor
 centered on the  
``magic $\gamma = 29.3$'';  the effect of the electric  field on the muon
 spin precession
cancels for muons at the magic momentum.     We point out the
relative sizes of the various terms in the spin equations,
 and show that for experiments that use the magic $\gamma$ and 
electric quadrupole focusing
to store the muon beam, any proposed effect that multiplies either
 the motional magnetic
 field
$\vec \beta \times \vec E$ or the muon pitching 
motion $\vec \beta \cdot \vec B$ term, will be smaller by three
or more orders of magnitude, relative to the spin precession due to the
storage ring magnetic field. 
We use a recently proposed General Relativity
correction~\cite{preprint} as an example,
to demonstrate the smallness of any such
contribution, and point out that their revised preprint~\cite{preprint2} 
 still contains a
conceptual error, that significantly overestimates the magnitude of 
their proposed correction. 
We have prepared this document in the hope that future authors will find it
useful,  should they wish to propose
corrections from some additional term added 
 to the Thomas equation, Eq.~\ref{eq:fullfreq}, below.  Our goal is
to clarify how the experiment is done, and how the
small corrections due to the presence of the
 radial electric field and  the vertical pitching motion of the
muons (betatron motion) in the storage ring are taken into account.

\end{abstract}

}

\large

\section{Introduction}

%
In a recent preprint~\cite{preprint}, Morishima et al., calculated a potential
general relativity (GR) effect 
 on the frequency of muon spin precession in a magnetic field. Several
 authors~\cite{Visser,Guzowski,Nikolic,Laszlo:2018,Venhoek:2018}, 
have posted papers on arXiv,
  questioning various aspects of these calculations, and others have 
done so in private communications to us.  More recently  Morishima et al.
updated their paper~\cite{preprint2}.
It is clear in both papers that these authors have misinterpreted the
 subtleties of the 
experimental technique, which if properly understood, would have prevented
them from claiming such a large correction to the measured muon spin
precession in their original paper~\cite{preprint}, which is perpetuated in
the update~\cite{preprint2}.

In this paper we examine the spin precession formulae, and calculate the
magnitude of the electric field term for the Brookhaven National
 Laboratory (BNL) E821
experiment.  To simplify this narrative, we have relegated
details of the beam dynamics in the storage ring to appendices. The appendices
include the derivation of 
corrections to the spin precession frequency from the electric field used to
provide vertical focusing, along with the derivation of the correction of the
vertical pitching motion of the beam, to appendices. 
These two effects give rise to  $\simeq 0.5$~ ppm and $\simeq 0.3$~ppm 
corrections respectively.

First we briefly put the physics motivation of, and the results from
 the Brookhaven Muon \g2 experiment in context,
and then explain how the assumptions in Refs.~\cite{preprint,preprint2}, or
any other ppm or less effect,  are not a relevant
concern for the interpretation of the BNL E821 results, or from the ongoing
muon $(g-2)$ experiment, E989 at Fermilab.

A spin 1/2 lepton  ($\ell = e,\,\mu,\,\tau$) has an intrinsic
 magnetic moment due to  its spin, given by the relationship
 \be \vec \mu_{\ell} =g_{\ell}
\frac{Qe}{2m_{\ell}}\vec{s}\,,\qquad g_{\ell} =2(1+a_\ell), \qquad a_\ell = \frac{g_\ell -2}{2} \, ,
\label{eq:anomaly}
\ee 
where $Q = \pm 1$, $e>0 $ and $m_\ell$ is the lepton mass.  
Dirac theory predicts
that $g \equiv 2$, but experimentally, it is known to be greater than 2.
This deviation from $g=2$, $a_\ell$ in Eq.~\ref{eq:anomaly},
 is the magnetic anomaly, which arises from radiative corrections
 (quantum fluctuations).

\begin{figure}[h!]
\begin{center}
  \includegraphics[width=0.95\textwidth,angle=0]{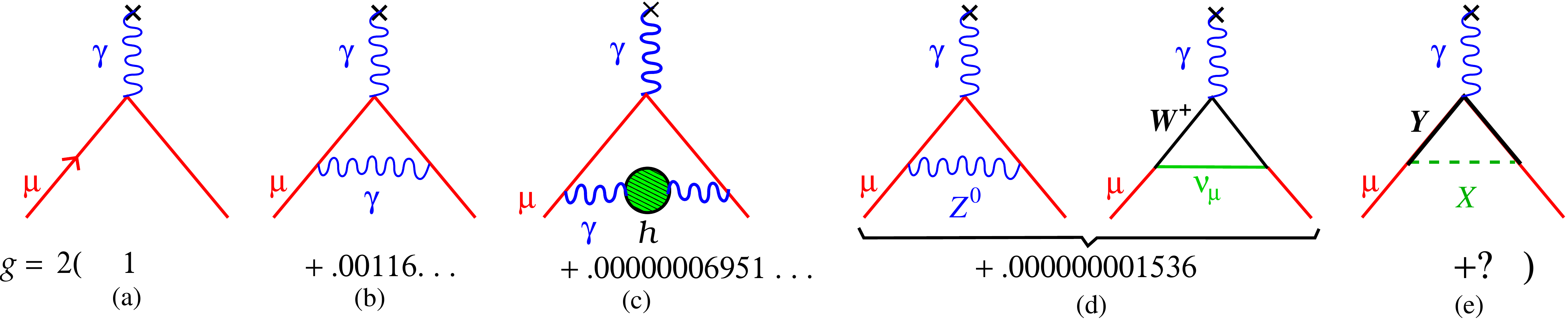}
  \caption{The Feynman graphs showing contributions to $g$ for each of the
  Standard Model forces, ordered by size: 
(a) The Dirac interaction.
 (b)  The lowest-order QED term $\alpha/2\pi$,
 which dominates the value of the anomaly.
 (c)  The hadronic vacuum polarization
contribution. (d) The lowest-order electroweak contributions. (The one-loop
 Higgs contribution is negligible.) (e) Potential contribution from new BSM
 particles $X$ and $Y$.
  \label{fg:contributions}}
\end{center}
\end{figure}

 In the Standard Model, $a_\mu$ gets measurable radiative
 contributions from QED, the strong interaction, and
from the electroweak 
interaction~\cite{Schwinger:1948,Aoyama1:2012,KNT,DHMZ,Jegerlehner:2017lbd,Gnendiger:2013},
\be
a^{SM} = a^{QED} + a^{Had} + a^{Weak} \, ,
\ee
which are shown diagrammatically in Fig.~\ref{fg:contributions} along with the 
 magnitudes of these contributions. 

The Standard Model value, $a_\mu^{SM}$, has
 been theoretically calculated with an uncertainty of about 
$\pm 0.3$~ppm~\cite{KNT,DHMZ,Jegerlehner:2017lbd}.
The largest contribution comes from the mass-independent
 single-loop (Schwinger) diagram~\cite{Schwinger:1948}, shown in
Fig.~\ref{fg:contributions}(b)\,\footnote{With his
  famous calculation  
that obtained  $a = (\alpha/2 \pi) = 0 .00116\cdots$,
Schwinger started an ``industry'',
which required  Aoyama, Hayakawa, Kinoshita and Nio
to calculate more than 12,672 diagrams to evaluate the tenth-order
(five loop) contribution~\cite{Aoyama1:2012}.  }. 
Should the experimental value~\cite{Bennett:2006},
\be
a_\mu^{E821} = 116\,592\, 091 (63) \times 10^{-11}
\ee
 differ from the Standard Model value at a
statistically significant level, it would reflect additional contributions
from as yet undiscovered particles beyond those of the Standard 
Model~\cite{Miller:2012,Stoeckinger2010}.  Using the hadronic contributions
from Refs.~\cite{KNT,DHMZ,Jegerlehner:2017lbd}, the difference between
experiment and the Standard Model theory
 is positive, with a statistical significance between
3.5 and 4 standard deviations.

\section{Past Experiments and the Spin Equations}

\subsection{The Basics}

When placed in a magnetic field at rest, the muon undergoes Larmor precession, 
\be
\vec \omega_{L_\mu} =- g_\mu \frac{Qe}{2m}\vec B \, ,
\ee
where $e$ is the magnitude of the electron charge, $Q=\pm 1$ and $m$ is the
muon mass. 
The very first measurements of the muon magnetic moment  were done
 at rest~\cite{garwin1,Cassels,garwin2}.  The first two determined that
$g$ was consistent with $2$.
The more precise third experiment
 demonstrated that $a_{\mu}$ was consistent with $\alpha/2\pi$,
demonstrating that a muon behaved like an electron when placed in a magnetic
field. The paper by Cassels et al.~\cite{Cassels} pointed out that if the
experiment were to be done with muons in flight, 
the difference between the spin
precession 
frequency and the cyclotron frequency would depend directly on the radiative
corrections, rather than on $g$.  For this reason, all subsequent experiments
were done with in-flight muons in a magnetic field, in order
to measure $a_\mu$ directly.

In the simplest case, in the absence of an electric field and when
the muon velocity
is perpendicular to a uniform magnetic field,
the rate at which the spin turns relative to the
momentum is given by the difference between the spin rotation frequency
$\omega_S$ and the cyclotron frequency $\omega_C$,
\begin{equation}
\vec \omega_{a_\mu} = \vec \omega_S - \vec \omega_C =-g_\mu\frac{Qe \vec B}{2m} -
(1-\gamma)\frac{Qe \vec B}{\gamma m}+ \frac{Qe \vec B}{\gamma m}=
 - \left(\frac{g_\mu-2}{2}\right) \frac{Qe}{m} \vec B
= - a_\mu\frac{Qe}{m} \vec B \, ,
\label{eq:diffreq}
\end{equation}
and is directly proportional to $a_\mu$.

\subsubsection{Determination of $a_\mu$ from $\omega_{a_\mu}$}

The experimental value of $a_{\mu}$ is derived from 
\be
\omega_{a_\mu}= \frac{e}{ m_{\mu}}a_{\mu} B= \frac{2a_{\mu}\omega_{L_\mu}}{ g_\mu}
=\frac{a_{\mu}\omega_{L_\mu}}{ 1+a_{\mu}},
\ee
with the appropriate small corrections such as the electric field and pitch
corrections described in the appendices. This can be written as
\be
a_{\mu}=\frac{\omega_{a_\mu}}{ \omega_{L_\mu}-\omega_{a_\mu}}\, ,
\label{eq:amu1}
\ee
where we have used 
\be
 a_{\mu}=\frac{g_\mu-2}{ 2}, \ 
 \omega_{L_\mu}= g_\mu\frac{e}{2m_{\mu}} B,\ \ 
       \ \  \hbar \omega_{L_\mu} =2 | \mu_\mu| B\, .
\ee
and $B =|B|$.
  
The \g2 experiments measure two frequencies,  
$\omega_{a_\mu}$ and $\omega_{L_p}$, the
latter being the Larmor frequency of protons in nuclear magnetic resonance 
(NMR) probes 
used to monitor the magnetic field, which are calibrated to the
Larmor frequency of a free proton.  
\be
\omega_{L_p} =  g_p\frac{e}{ 2m_{p}}B\ \ {\rm and}\ \ \hbar
\omega_{L_p} = 2 \mu_p B\, .
\ee
 Dividing the
numerator and denominator in Eq.~\ref{eq:amu1} by $\omega_{L_p}$, we get
\be
a_\mu = \frac{{\omega_{a_\mu}}/{\omega_{L_p}}}
 {|{\mu_\mu}|/{\mu_p}- 
{\omega_{a_\mu}}/{\omega_{L_p}}} =\frac{R}{\lambda- R}\, ,
\ee
where $R= \omega_{a_\mu}/\omega_{L_p}$ was measured experimentally at
BNL by E821~\cite{Bennett:2006},  and 
$\lambda = \mu_\mu^+/\mu_p = 3.183\,345\,142 (71)$~\cite{CODATA},
determined from muonium hyperfine structure splitting~\cite{lambda-fc}.  
This equation
was used to determine the value of $a_\mu$ in Brookhaven E821, which required
the assumption of CPT invariance to determine $a_{\mu^-}$~\cite{Bennett:2006}.

Fermilab 
 E989 proposes to use the equivalent combination 
of constants~\cite{TDR} rather than $\lambda$,
\be
a_\mu = \left( \frac{g_e}{2}\right) 
\left( \frac{m_\mu}{m_e}\right)
\left( \frac{\mu_p}{\mu_e}\right)
\left( \frac{\omega_{a_\mu}}{ \omega_{L_p}}\right) \, ,
\label{eq:a_mu_fc}
\ee
 to determine $a_\mu$ from the two measured frequencies.
 The additional ratios 
appearing in Eq.~\ref{eq:a_mu_fc} are well known from other experiments: 
$g_{e}/2=1.001\,159\,652\,180\,73(28)$ (0.28 ppt)~\cite{gabrielse-fc},  
$m_{\mu}/m_{e}=206.768\,2826(52)$ (22 ppb)~\cite{CODATA,PDG} and
${\mu_p}/{\mu_e}=658.210\,6866(20)$ (3 ppb)~\cite{CODATA}.

\subsection{Vertical Focusing with an Electric Field}

In a real experiment, the magnetic field that appears in Eq.~\ref{eq:diffreq}
 must be averaged over the muon ensemble, which consists of a range of momenta
and positions in the storage ring.  This averaging requires
 detailed information on
 the muon orbits, and the value of the magnetic field at each location
 ($x,y,z$) in the muon orbits.  Of three \g2
 experiments at CERN~\cite{cern1,cern1a,cern2}, 
the first two used magnetic gradients to contain the
 muons, the second being a storage ring
 experiment.   However, to go beyond the precision
 of the second  CERN experiment, it became necessary 
to find a different way to provide
 vertical focusing for the stored muon beam, since the presence of gradients
 in the magnetic field made it difficult to know the magnetic field
averaged over the muon beam distribution to the parts per million (ppm) level. 

To overcome this issue, the third CERN collaboration~\cite{cern3a,cern3b} 
realized that one could employ a uniform magnetic
dipole field and in addition   an electrostatic quadrupole field for vertical
focusing\,\footnote{This arrangement of magnetic and electric fields is 
  essentially a Penning trap.}.  To a relativistic particle, 
 an electric field is perceived to be a
combination of electric and magnetic fields.  The resulting motional magnetic
field (MMF) through the  $\vec \beta \times \vec E$ term
can cause the spin to precess.
For the more general case where $\vec \beta \cdot \vec B \neq 0$ and
$\vec E \neq 0$, the cyclotron
rotation frequency becomes:
 \be
\vec  \omega_C =- \frac{Qe}{m}\left[ { \vec B \over \gamma} 
- {\gamma \over \gamma^2 -1} 
\left({ \vec \beta \times \vec E \over c}\right) \right],
\label{eq:cyc-E}
\ee
 and the spin rotation frequency becomes
\be
\hspace*{-1.0cm} \vec \omega_S = - \frac{Qe}{ m } \left[
\left({g \over 2} -1 + {1\over \gamma}\right) \vec B
- \left( {g \over 2} -1 \right){\gamma \over \gamma + 1}(\vec \beta \cdot
\vec B)\vec \beta -
\left( {g \over 2} - {\gamma \over \gamma + 1}\right) 
\left( { \vec \beta \times \vec E \over c}\right )
\right]\,.
\label{eq:fullfreq}
\ee
This equation was first discovered by L.H. Thomas in
1927~\cite{Thomas:1927}.   We use the form given in Eq.~11.170 in Jackson's 
text~\cite{Jackson:1990},   which is equivalent to Thomas' Eq. 4.121 in 
Ref.~\cite{Thomas:1927}, but in modern notation~\footnote{Bargmann, Michel and 
Telegdi~\cite{BMT:1959} also studied this
problem.}.

Using  $a_{\mu} = (g_{\mu} -2)/2$, we find that
the spin difference frequency is\footnote{Strictly speaking, the rate of change of the angle between the spin and the
momentum vectors,
$|\vec\omega_{a_\mu}|$= `precession frequency',
is equal to $|\vec\omega_{diff}|$ only if
$\vec\omega_S$
and $\vec \omega_C$ are parallel. For the E821 and E989 experiments,
the angle between $\vec\omega_S$ and $\vec\omega_C$ is always small and the
rate of oscillation of $\vec\beta$ out of pure circular motion is fast compared
to $\omega_{a_\mu}$, allowing us in the following discussion
the make the approximation that $\vec\omega_{a_\mu}\simeq\vec\omega_{diff}$.
More general calculations, where this approximation is not made, are found in
References~\cite{Farley:1972,Combley:1974,Field:1974,Farley:1990}. 
For the E821 and E989 experimental conditions, the results
presented here are the same as those in these references. }
\begin{equation}
\vec \omega_{diff}=  \vec \omega_S -\vec\omega_C     \simeq \omega_{a_\mu} 
=  - \frac{Qe}{ m}
\left[ a_{\mu} \vec B -  
a_{\mu}\left( {\gamma \over \gamma + 1}\right)
(\vec \beta \cdot \vec B)\vec \beta 
- \left( a_{\mu}- {1 \over \gamma^2 - 1} \right) 
{ {\vec \beta \times \vec E }\over c }\right]\,.
\label{eq:Ediffreq}
\end{equation}

With the presence of the electric field in the third term,
 and the fact that the velocity is
not perpendicular to the magnetic field in the second term,
 small corrections must be 
accounted for in the determination of the muon anomaly, which are
derived below in Appendices II and III.  

\subsection{The Magic $\gamma$ and  a correction
 to the $\vec \beta \times \vec E$ term}

The discussion in Ref.~\cite{preprint} is centered around the 
spin Eq.~\ref{eq:Ediffreq} above, which is used to describe the
muon spin motion \underbar{relative} to the momentum vector
 in the Brookhaven muon \g2 storage ring~\cite{Danby:2001,Semertzidis:2003}.
Under the approximation that the muon velocity is perpendicular to the magnetic
field ($\vec \beta \cdot \vec B = 0$),  Eq.~\ref{eq:Ediffreq} reduces to
\be
\vec \omega_{a}
=  - \frac{Qe}{ m}
\left[ a_{\mu} \vec B -
 \left( a_{\mu}- {1 \over \gamma^2 - 1} \right) 
{ {\vec \beta \times \vec E }\over c }\right]\, .
\label{eq:magicgam}
\ee

\begin{figure}[h!]
\begin{center}
\includegraphics[width=0.3\textwidth,angle=-90]{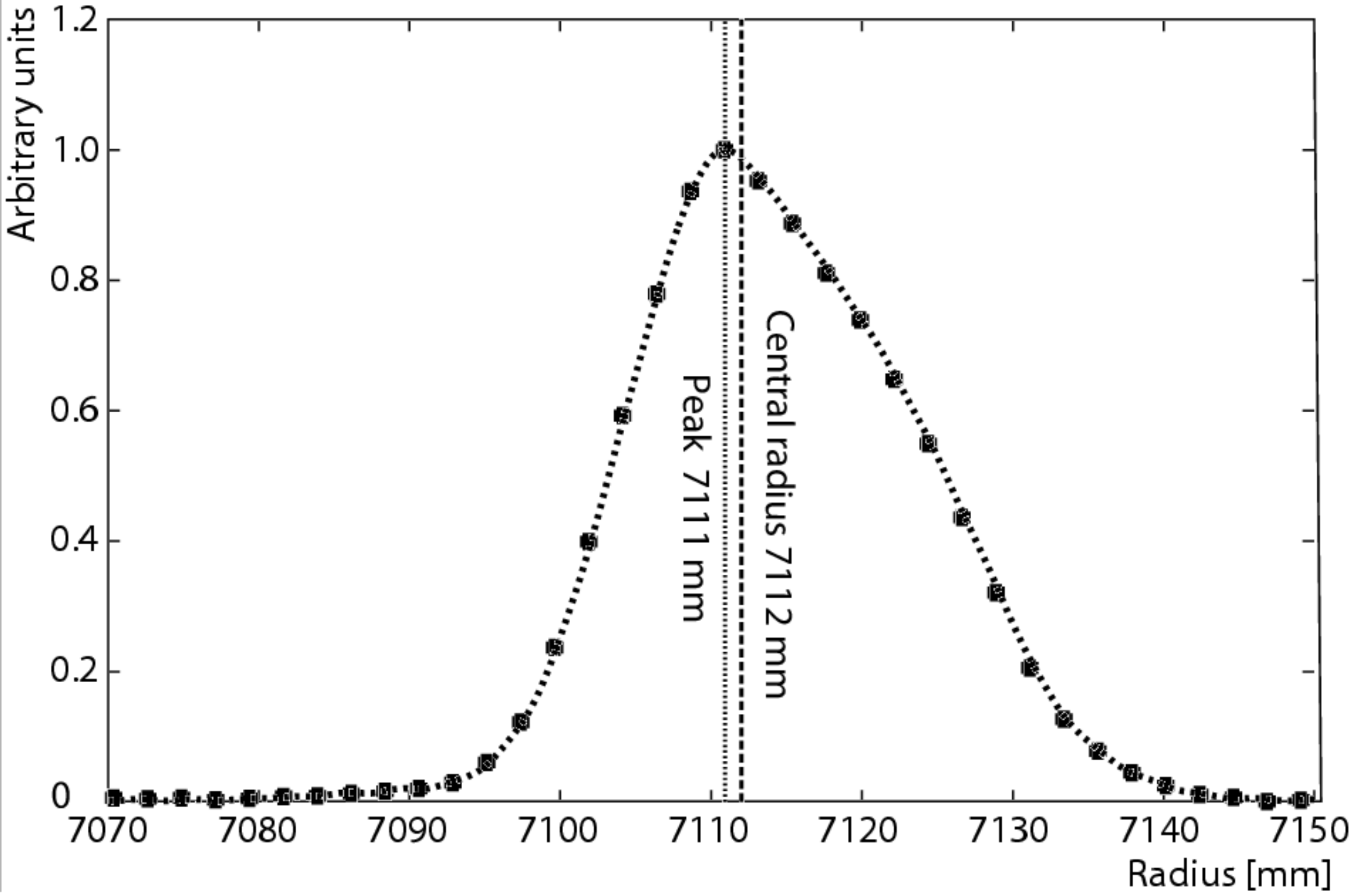} 
  \caption{{The measured distribution of equilibrium radii in the \g2 storage
    ring.  This distribution was used to determine the radial electric field
    correction that is derived in Appendix II. (From Bennett
    et. al,~\cite{Bennett:2006}.) }  
  \label{fg:eq-radii}}
\end{center}
\end{figure}

If the value of the relativistic $\gamma$-factor is chosen to be
$\gamma_m = 29.304$, $p_m=3.09$~GeV/c, then 
\be
\left( a_{\mu}- {1 \over \gamma^2 - 1} \right) =0 \, ,
\ee
and the $\vec \beta \times \vec E$ term does not contribute
 to the muon spin precession. 
However, a well understood correction must be made to the measured spin
frequency to account for the fact that not all muons are at the magic
$\gamma$ (see Appendix II). In the storage ring of the E821 and E989
experiments, there is an ensemble of muons with a range of momenta determined
by the momentum acceptance of the storage ring, which is $\pm
0.5\%$. Therefore, with the muons centered about the magic $\gamma$, the
maximum 
$\gamma$ range of the stored muons is $\gamma = \gamma_m \pm 0.5\%$. Due to
the phase space acceptance of the storage ring, the
distribution of momenta is approximately Gaussian with a width, 
$\sigma \simeq 0.15\%$, as shown 
in Fig.~\ref{fg:eq-radii}.

Suppose that an effect $\xi$, such as that
proposed in Ref.~\cite{preprint}, multiplies
the  $\vec \beta \times \vec E$ term in Eq.~\ref{eq:Ediffreq}.  The question
is: How would this term $\xi$ change the spin difference
precession frequency?   To get an
order of magnitude estimate we consider the term:  
\bea
&  &    \left[ 
\left(
 a_{\mu}- {1 \over \gamma^2 - 1} 
\right) 
\pm \xi 
\right]
\frac{\vec \beta \times \vec E}{c} \label{eq:xi} \\
&=& 
\underbrace{\left( a_{\mu}- \frac{1}{\gamma^2 - 1} \right)
\frac{\vec \beta  \times \vec E}{c}}_{{\rm radial} \ E{\rm- field \ correction}}
\  \pm\  \xi 
\frac{\vec \beta  \times \vec E}{c} \label{eq:xiuse}
\eea
The reason that we separate these two effects is because the first term in
Eq.~\ref{eq:xiuse} is used to calculate
the radial electric field  correction as we explain in Appendix II, and the
effect of the $\xi$ term must be calculated separately.  
So any physical effect that
would change the muon spin difference frequency must be multiplied
\underbar{separately}  by the
value of $(\vec \beta  \times \vec E)/{c}$  as indicated in Eq.~\ref{eq:xiuse}.
 This is an
important point, which Morishima et al. missed, as they worried about the 
magic $\gamma$ cancellation for each muon in the beam\,\footnote{This
  misunderstanding became clear in a discussion with two of the authors while
  BLR was visiting the University of Nagoya on February 16, 2018.
  Unfortunately,
their revised preprint~\cite{preprint2} still contains this conceptual error.}.
We use the proposed correction of Morishima et al.~\cite{preprint}, where
$\xi =  -  2.8 \times 10^{-9}$, as an example
\footnote{Their proposed
  correction of $ 2.8 \times 10^{-9}$
 is roughly twice the electroweak contribution~\cite{Gnendiger:2013} to \amu
 of \break  $1.54\times 10^{-9}\  (1.31)$~ppm .},
 and we leave 
 the the correctness of their General Relativity derivation
for other authors to discuss.

\subsection{Values of the E821 pitch and radial $\vec E$ field
  corrections}

Before we calculate the value of $(\vec \beta  \times \vec E)/{c}$ in
Section~\ref{sct:OOM},  we present the 
corrections made to the E821 data for the vertical pitching of the beam 
(the $ \vec \beta \cdot \vec B$ term),
and for the radial electric field ($\vec \beta \times \vec E$) term. 
See Ref.~\cite{Bennett:2006} and the appendices below for more details.
In each case, the correction requires that we increase the measured
value of $\omega_a$.
For the 2001 data collection period~\cite{Bennett:2006},
where data were collected at two different $n$ values, the electric field
corrections were $C_E(\rm{low}-n) = 0.47 \pm 0.054$~ppm and
$C_E(\rm{high}-n) = 0.50 \pm 0.054$~ppm.  The pitch corrections  were 
$C_P (\rm{low}-n) = 0.27 \pm 0.036$~ppm and
 $C_P (\rm{high}-n) = 0.32 \pm0.036$~ppm. These corrections were the 
\underbar{only}
 corrections made to the muon spin precession frequency in E821, and their
 magnitude demonstrates that any additional effects that multiply 
$\vec \beta \cdot  \vec B$ or $\vec \beta \times \vec E$ should be expected
to be small.


\section{Order of magnitude of the effect of $\xi$ on $\omega_{a_\mu}$
\label{sct:OOM}} 

\subsection{Correction from the Motional Magnetic field}

An order of magnitude calculation, demonstrates that a
$\xi = 2.8\times 10^{-9}$ contribution to 
the motional magnetic field
term ($\vec \beta \times \vec E$ term) makes a negligible contribution
to the measurement
of $ a_{\mu}$ relative to the 0.5~ppm (BNL), and 0.14~ppm
(projected Fermilab sensitivity) levels of precision.  

The principal feature that must be recognized when calculating a new effect
on the spin difference  frequency,  is that $\omega_{a_\mu}$ is dominated by
the $a_\mu B$ term.   Any quantity that multiplies the 
$\vec \beta \cdot  \vec B$
or $\vec \beta \times \vec E$ term will be small, compared to the spin
precession caused directly by the 1.45 T storage ring magnetic field,
\be
 - \frac{Qe}{ m}
 a_{\mu} \vec B \, .
\label{eq:dominant}
\ee

For the correction reported in Ref.~\cite{preprint}, 
where $\xi =-2.8 \times 10^{-9}$, the shift in $\omega_{a_\mu}$ would be
\be
-2.8 \times 10^{-9}\left\langle  \frac{\vec \beta \times \vec E}{
    c}\right\rangle_{avg}\, ,
\ee
where we have ignored for the moment the common factor $Qe/m$.
The average refers to an average over the distribution of
stored muons.
This is the term that Refs.~\cite{preprint,preprint2}
 claim causes a sizable shift in the
value of $a_{\mu}$.
We now show in detail that their conclusion is incorrect, 
and in fact that this term is negligibly small compared to the dominant
contribution to $\omega_a$, which comes from the term in Eq.~\ref{eq:dominant}.

The ``magic momentum'' is  $p_m\simeq 3.09$ GeV/c.
The distribution of momenta in the storage ring is approximately Gaussian
with $\sigma \simeq 0.15\%$. 
With that range of momenta the $E$-field correction
is about 0.5 ppm, which in E821 was known to $\simeq 10$\%.
 We do not learn anything with any accuracy on $a_{\mu}$ by tuning
for the magic momentum. Nearly all the information on $a_{\mu}$ is derived
from the value of $\omega_{a_\mu}$, since the two are proportional to each other,
due to the \underbar{dominance} of the $B$-field term on the spin precession
frequency. The value of 
$\omega_{a_\mu}$ is completely insensitive to  the magic $\gamma$ used in the
$\vec \beta \times \vec E$, except
 at the fraction of a ppm level.
Therefore we derive no information on $a_{\mu}$ from the $E$-field term because
of the smallness of the $\vec \beta \times \vec E/c$ term. For
example, tuning the momentum until the coefficient of the $E$-field term is
zero will tell us nothing about the value of $a_{\mu}$ at the desired ppm
sensitivity. In fact, because of the width of the
 momentum distribution, which is
significantly larger than the ppm level, one has to treat the ensemble of
muons as a whole, with the correction given in Appendix II.

To calculate the magnitude of the frequency change caused by the term 
 $\xi$ in Eq.~\ref{eq:xiuse} has on the muon spin
precession,
 we need to evaluate
$\vec \beta\times \vec E$, \underbar{averaged} over the population 
of stored muons.
In the muon storage ring, the muon velocity (0.9994 c) is dominantly along
the ring azimuth. We only need to consider 
the radial component of $\vec E$, since $\vec \beta \times \vec E$ is directed
along the $B$-field and therefore is the only component that contributes to the
precession frequency, 
$\omega_{a_\mu}$, to first order.

The electric potential for a quadrupole electric field, with the poles
situated along the radial and vertical ($B$-field) axes, is 
\be
V=-{1\over 2}\kappa (x^2-y^2)
\ee
where $\kappa$ is the electric field gradient, $x=r-r_0$,  $r$ is the
radius of the muon trajectory perpendicular to the $B$-field, and $r_0=7.112$~m
is the central radial position 
in the storage region.
The radial field is given by 
\be
E_r\simeq E_x=-{\partial V\over \partial x}=\kappa x.
\ee
The electric
quadrupoles are split into four equal segments
azimuthally around the storage ring, 
covering about 43\% of the ring. 

For the E821 experiment, for positive muon storage,
typically $V\simeq -25,000$~V at the horizontal
electrodes located at $(r_0-0.05)~{\rm m}$ and $(r_0+0.05)~{\rm m}$.
Similarly the vertical electrodes have
$V=+25,000$ V at $y=0.05$ m and $y=-0.05$ m. Thus the electric field gradient
is 
\be
\kappa = {2\times 25000\ {\rm V} \over (0.05)^2 {\rm m}^2}
=2 \times 10^{7}\ {\rm V/m^2}.
\ee

The typical radial electric field as a function of $x$ is therefore
\be
E_r=E_x=-{\partial V\over \partial x}=\kappa x=2\times 10^7~x~{\rm V/m}\, .
\ee
The average value of $\vec \beta \times \vec E$ is simply $\beta=0.9994$ times
the average value of $E_r$. The average radial position of the stored muon
beam in E821 is about 7.115 m, or 0.003 m larger than the central ring radius.
The average value of $E_r$ is reduced by a factor of 0.43
to account for the partial coverage of electric quadrupoles inside the ring,
thus the average value
\be
(E_r)_{avg}=0.43\times 0.003\times 2\times 10^7=2.6\times 10^4~{\rm V/m}.
\ee

The relative contribution of $\xi =  -  2.8 \times 10^{-9}$
 to the precession frequency,
${\Delta \omega_{a_\mu}/ \omega_{a_\mu}}$, which is very nearly
${\Delta a_{\mu}/ a_{\mu}}$, 
is thus given by dividing the $\xi$ contribution to the $E$-field term, by the
much larger $B$-field term, Eq.~\ref{eq:dominant}:
\be
\frac{\Delta a_{\mu}}{ a_{\mu}}={\xi \times \beta 
\times (E_r)_{avg}\over cBa_{\mu}}
\ee
To calculate a numerical value, we use
$ B=1.45$~T, $\beta=0.9994$, $a_{\mu}=1.1659\times 10^{-3}$.
Using these values and  $\xi =2.8\times 10^{-9}$ 
one gets: 
\be 
\frac{\Delta a_{\mu}}{ a_{\mu}}  =1.4 \times 10^{-10}\, ,
\ee
or about 0.14 ppb.

We have shown that the proposed GR contribution of Eq.~40 of
Ref.~\cite{preprint}, and in section 3.4 of Ref.~\cite{preprint2}
 when applied correctly to the storage-ring
 experiments with electric quadrupole focusing,
 is completely negligible when 
compared to the experimental uncertainty of 0.54~ppm of BNL E821,
or to the uncertainty of 0.140~ppm expected at Fermilab.
 In order to have an observable
effect on the determination of \amu, one must have
 $\xi \geq 1 \times 10^{-6}$.  
Thus the conclusions in section 3 of Ref.~\cite{preprint}
and section 3.4 of Ref.~\cite{preprint2} 
are incorrect.


\subsection{Correction from the $\vec \beta \cdot \vec B$ term}

We now briefly consider the vertical pitching motion, which introduces a term
proportional to $\vec\beta\dot \vec B$,
\be
\vec\omega_{a_\mu}\simeq\vec \omega_{diff}
=   - \frac{Qe}{ m}
\left[ a_{\mu} \vec B -  a_{\mu}\left( {\gamma \over \gamma + 1}\right)
(\vec \beta \cdot \vec B)\vec \beta 
 \right]
\ee
into the spin difference equation.  As mentioned above this term caused a
small change of $\simeq 0.3$~ppm correction to the precession frequency,
which is derived in Appendix III.  The situation is similar to the
motional magnetic field correction, {\it viz.} an additional term would also
multiply a small number that is on the order of 0.3~ppm, and if it is the
same order of magnitude of $\xi$ discussed above would have a negligible
effect on the spin precession.

\section{Summary}

We have outlined the approximate expressions for spin precession used to
derive $a_{\mu}$ from experimental data in BNL E821. We have included
discussion of corrections associated with the focusing electric field and the
betatron motion of muons in the storage ring.
We have shown that
any small effect $\xi$, perhaps from
 general relativity or from some other source,  that multiplies the
$(\vec \beta \times \vec E)$ term in the spin equations
will not contribute to the muon spin rotation at a measurable level,
unless it is greater than $10^{-6}$.  For larger values, the exact details
would need to be understood, in order to calculate the effect on the \g2
measurement.  Thus the
$\xi = 2.8 \times 10^{-9}$ contribution proposed by Morishima, et
al.~\cite{preprint,preprint2}
 does not contribute at a measurable
level to the muon \g2 experiments,
 since when calculated correctly, its effect on the measurement
 is at the tenths of a part per billion level, three orders of
magnitude smaller than the expected experimental precision of the
 Fermilab E989 final result. Therefore, the conclusions in \textsection 3 of
 Ref.~\cite{preprint}, and conclusions in \textsection 4.3, including  Eq.~77
 of Ref.~\cite{preprint2} are
 incorrect.  The authors have misinterpreted
the means by which the \g2  experiments have determined $a_\mu$, and have
consequently greatly overestimated the size of the correction based on their
proposed GR term.

\section{Acknowledgments}

We wish to thank  Gianguido Dall'Agata, Alexander
Keshavarzi and  Massimo Passera for useful comments
 on this manuscript, and our g-2
collaborators for helpful conversations. This work was supported in part by
the U.S. Department of Energy, Office of High Energy Physics.

\newpage

\appendix{\noindent \bf{Appendix 1: Beam dynamics in the \g2 storage ring.}}\\

\begin{figure}[h!]
\begin{center}
  \includegraphics[width=0.8\textwidth,angle=0]{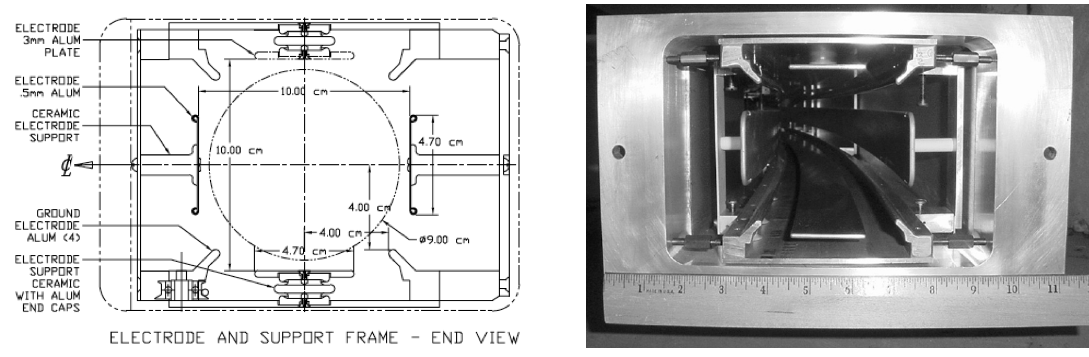}
  \caption{Left: A drawing of the electrostatic quadrupoles. Right: A
    photograph of the quadrupoles. 
 From Semertzidis, et al.,~\cite{Semertzidis:2003}.
  \label{fg:quads}}
\end{center}
\end{figure}

In E821 at BNL, and in the upcoming experiment at Fermilab, muons are stored
in a magnetic storage ring~\cite{Danby:2001} with a uniform magnetic field
and   electrostatic 
quadrupoles~\cite{Semertzidis:2003}, shown in Fig.~\ref{fg:quads},
 to provide vertical focusing.
With the weak focusing electric quadrupoles, the 
 \g2 storage ring acts like a weak-focusing betatron.~\cite{TDR}.  
 The muons execute
harmonic motion in both the vertical and horizontal directions as they go
around the storage ring.

An important quantity of the storage ring is the field index $n$ which
is given by:
\begin{equation}
 n=\frac{\kappa R_0}{ v B_0}
\label{eq:field-index}
\end{equation}
where $\kappa$ is the quadrupole gradient strength.  
The frequencies of the radial ($x$) and vertical ($y$)
 betatron motions are determined by the field index:
\begin{equation}
 f_x = f_C \sqrt{1-n}\simeq 0.929 f_C \quad {\rm and} \quad
f_y = f_C \sqrt{n} \simeq 0.37 f_C,
\label{eq:betafreq}
\end{equation}
where $f_C = 6.7$~MHz is the cyclotron frequency. 
 A significant amount of the E821
data were taken with $n=0.137$, where the potentials on the quadrupole plates
were $\pm 25$~kV. 
The muon spin precession frequency is $f_a= 229.073$~kHz

The field index also determines the angular acceptance of the ring.  
The maximum horizontal and vertical angles of the muon momentum
are given by
\be
\theta^x_{\rm max} = \frac{ x_{\rm max} \sqrt{1-n}}{ R_0},
\quad {\rm and} \quad 
\theta^y_{\rm max} = \frac{ y_{\rm max} \sqrt{n}}{ R_0},
\label{eq:max-angle}
\ee
where $x_{\rm max},y_{\rm max} =  45$~mm is the radius of the storage
aperture, and $R_0 = 7.112$~m.
For a betatron amplitude $A_x$ or $A_y$ less than 45~mm, the 
maximum angle is reduced, as can be seen from the above equations.

\newpage

\appendix{\noindent{\bf Appendix II: The radial electric field correction 
\label{sct:E-cor}}\\

We consider the $\vec \beta \times \vec E$ term first. For muons not at the
magic radius (magic $\gamma$) we have the following correction. 
\be
\omega'_a = \omega_{a_\mu}\left[ 1 - \beta {E_r \over c B_y} 
\left( 1 - {1 \over a_{\mu} \beta^2\gamma^2 }\right) \right],
\ee
where $\omega_{a_\mu}=-a\frac{Qe}{m}B$.
Using $p = \beta \gamma m = (p_m + \Delta p)$. 
Note that the quantity $(E_r/c)B_y$ is the ratio of the motional magnetic field
to the 1.45~T storage ring field.

After some algebra one finds
\be
{\omega_{a_\mu}' - \omega_{a_\mu} \over \omega_{a_\mu}} = 
{\Delta \omega_{a_\mu} \over \omega_{a_\mu}} = -2 {\beta E_r \over c B_y }
\left({ \Delta p \over p_m}\right).
\ee
 Now
\be
{\Delta p \over p_m} = (1-n){\Delta R \over R_0} = (1-n) {x_e\over R_0},
\ee
where $x_e$ is the muon's equilibrium radius
of curvature  relative to the central orbit.
The mean electric quadrupole field experienced by the muon is
\be
\langle E_r\rangle = \kappa x = { n \beta c B_y \over R_0} x_e.
\ee
where $x_e$ is the equilibrium radius of the muon.
We obtain
\be 
{\Delta \omega \over \omega} = - 2n(1-n)\beta^2 {x x_e \over R^2_0},
\ee
so the effect of muons not at the magic momentum is to lower the observed
frequency.
For a quadrupole focusing field plus a uniform magnetic field,
the time average of $x$ is just $x_e$, the equilibrium radius, so 
the electric field correction is given by
\be
C_E = {\Delta \omega \over \omega} 
=  - 2n(1-n)\beta^2 {\langle x^2_e\rangle \over R^2_0 },
\ee
where $\langle {x^2_e}\rangle$
 is determined from experimental data (see Ref.~\cite{Bennett:2006})}

\newpage

\appendix{{\noindent \bf Appendix III: The pitch correction}}
\\

The ``pitch'' correction from the $\vec \beta \cdot \vec B \neq 0$ term also
contributes a small correction to the measured precession frequency. In this
derivation, we follow J. Paley~\cite{Paley:2004}.  More general
derivations can be found in Refs.~\cite{Farley:1972,Field:1974,Farley:1990}.
 
\begin{figure}[ht]
\centering
\includegraphics[width=0.4\textwidth,angle=0]{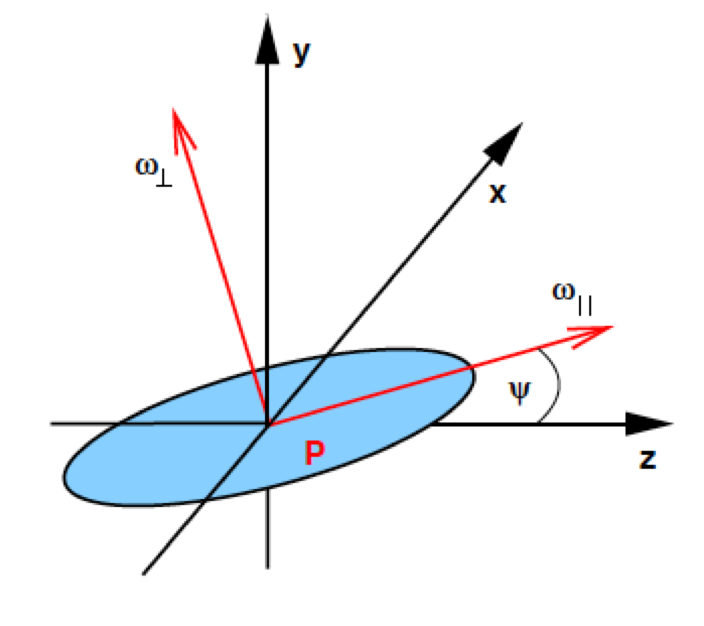}

  \caption{The coordinate system of the pitching muon.  The angle $\psi$
varies harmonically with the vertical betatron frequency. 
The vertical direction is $\hat y$ and
 $\hat z$ is the azimuthal (beam) direction.
       \label{fg:pitch-cs}}
\end{figure}

In the approximation that all muons are at the magic $\gamma$,
we set $a_{\mu}-1/(\gamma^2-1)=0$ in Equation~\ref{eq:Ediffreq} and obtain
\be
\vec\omega_{a_\mu}\simeq\vec \omega_{diff}
=   - \frac{Qe}{ m}
\left[ a_{\mu} \vec B -  a_{\mu}\left( {\gamma \over \gamma + 1}\right)
(\vec \beta \cdot \vec B)\vec \beta 
 \right].
\label{eq:diff-pitch}
\ee
The geometry is shown in Fig.~\ref{fg:pitch-cs}. 
The pitch angle $\psi= \psi_0 \cos\omega_{b_{y}}t $ 
oscillates with the vertical betatron frequency 
\be
\omega_{b_{y}} = 2 \pi \sqrt{n}f_C \simeq 2 \pi \times 2.5~{\rm MHz}
\ee
where $n$ is the field index from Appendix I, and $f_C$ is the cyclotron
frequency.  Thus $\omega_{b_{y}}$ is significantly larger than the spin
precession frequency $\omega_{a_\mu} \simeq 2 \pi \times 0.23~{\rm MHz}$, so we are
justified in averaging out the pitching motion.

We adopt the (rotating) coordinate system
shown in Figure~\ref{fg:pitch-cs}, where $\vec \beta$ lies in the $yz$-plane, 
$z$ being the direction of propagation, and $y$ being vertical in the storage
ring. The $x$ and $z$ axes rotate with the angular frequency 
\be
\omega = \frac{Qe}{m \gamma}B_y\, .
\ee
The frequency consists of two components, $\omega_\perp$ and
$\omega_\parallel$,
\be
\omega_\perp = \omega_{a_\mu} = \omega_y \cos \psi - \omega_z \sin \psi
\label{eq:wperp}
\ee

 Assuming $\vec B = \hat y B_y$, 
$\vec \beta = \hat z \beta_z +  \hat y \beta_y 
=\hat z \beta \cos \psi  + \hat y \beta \sin \psi $, 
\be
\vec \omega_{a_\mu}' = - \frac{Qe}{ m}
 [  a_{\mu}\hat y  B_y -  a_{\mu}\left( {\gamma \over \gamma + 1}\right)
 \beta_y B_y( \hat z \beta_z +\hat y \beta_y )].
\ee
we get
\be
\omega_{a_y} = -\frac{Qe}{m}a_\mu B_y \left[ 1 - 
\left( \frac{\gamma}{\gamma +1} \right)\beta_y^2   \right] 
= -\frac{Qe}{m}a_\mu B_y
\left[ 1 - \left(\frac{\gamma}{\gamma +1}\right)\beta^2
\frac{\beta^2_y}{\beta^2} \right]
\ee
Using 
\be
\frac{\beta_y}{\beta} = \sin \psi \simeq \psi\, , \ {\rm and }\ \ 
\frac{\gamma \beta^2}{ \gamma +1} = \frac {\gamma - 1}{\gamma}\,,
\ee
we get
\be
\omega_{a_y}=
 -\frac{Qe}{m}a_\mu B_y\left[ 1 - \left(\frac{\gamma -
       1}{\gamma}\right)\psi^2\right]
\label{eq:way}
\ee
For $\omega_{a_z}$ we have
\be
\omega_{a_z} = \frac{Qe}{m}a_\mu B_y
 \left(\frac{\gamma}{\gamma+1}\right)\beta_y\beta_z
= \omega_{a_z} = \frac{Qe}{m}a_\mu B_y
 \left(\frac{\gamma}{\gamma+1}\right)\beta^2\frac{\beta_y^2}{\beta^2}
\frac{\beta_z}{\beta_y}   \, .
\ee
Using $\beta_y/\beta_z =\tan\psi \simeq \psi$ 
\be
\omega_{a_z} = -\frac{Qe}{m}a_\mu B_y \left(\frac{\gamma
    -1}{\gamma}\right)\psi\,.
\label{eq:waz}
\ee
Substituting Eq.~\ref{eq:way} and Eq.~\ref{eq:waz} into Eq.~\ref{eq:wperp}
gives 
\be
\omega_{a_\mu} = \omega_\perp \simeq  -\frac{Qe}{m}a_\mu B_y
\left(1 - \frac{\psi^2}{2} \right)
=   - {Qe \over m} a_{\mu} B_y 
\left( 1 - {{\psi_0^2 cos^2 \omega_yt } \over 2}\right)
\ee
Taking the time average of the oscillating term we get
\be
  - {Qe \over m} a_{\mu} B_y 
\left( 1 - {{\psi_0^2 } \over 4}\right)
\ee
So the pitch correction is
\be
C_p = -  {\langle \psi_0^2\rangle \over 4}
=  - { n\over 2} {{ \langle y^2 \rangle }  \over R_0^2},
\ee
where we have used Equation~\ref{eq:max-angle},
$\langle \psi_0^2 \rangle = { { n \langle y^2 \rangle } / R_0^2}$, and
$\langle y^2\rangle = 0.5 \langle  y_0^2 \rangle$.
In E821, the quantity $ \langle y^2 \rangle $  was both 
determined experimentally and from
simulations.  
\newpage

\end{document}